\documentclass[prx,twocolumn,
]{revtex4}

\usepackage{graphicx}
\usepackage{amsmath}
\usepackage{amssymb}
\usepackage{amsfonts}
\usepackage{color}
\usepackage{float} 
\usepackage{siunitx} 

\newcommand{\subs}[1]{\ensuremath{_{\text{#1}}}}
\newcommand{\Ohm}{\ensuremath{\Omega}}

\definecolor{figorange}{RGB}{247,113,30}
\definecolor{figred}{RGB}{211,22,45}
\definecolor{figblue}{RGB}{41,171,226}
\definecolor{figgreen}{RGB}{0,146,69}

\begin{document}

\title{A Magnetic Persistent Current Switch at milliKelvin Temperatures}
\date{\today}
\author{Bob van Waarde} 
\email{waarde@physics.leidenuniv.nl}
\affiliation{Kamerlingh Onnes Laboratory, Leiden University, PO Box 9504, 2300 CA, Leiden, the Netherlands} 
\author{Olaf Benningshof} 
\affiliation{Kamerlingh Onnes Laboratory, Leiden University, PO Box 9504, 2300 CA, Leiden, the Netherlands} 
\author{Tjerk Oosterkamp} 
\affiliation{Kamerlingh Onnes Laboratory, Leiden University, PO Box 9504, 2300 CA, Leiden, the Netherlands} 

\begin{abstract}
We report the development of a magnetically driven Persistent Current Switch operated in a dilution refrigerator.
We show that it can be safely used to charge a $60$ mH coil with $0.5$ A at $11$ mK, 
which heats up the dilution refrigerator to $60.5$ mK.
Measurements at $4$ K on a $440$ \si{\micro\henry} coil 
reveal a residual resistance of $R \leqslant 3.3$ p$\Ohm$.
\end{abstract}

\maketitle


\section{Introduction}

There are many applications
in which a stable, low noise magnetic field is desired,
such as in MRI, NMR or qubit studies.
An elegant way of establishing such a magnetic field is by the use of a Persistent Current Switch (PCS).
In a PCS a superconducting coil is shunted by a superconducting shortcut such that together they form a closed resistanceless circuit.
The coil can be charged by briefly switching the shortcut to the resistive state.
The circuit then becomes an $RL$-circuit, and a power source can be connected to inject a current into the coil.
Back in the superconducting state, the flowing current is in principle stable and low-noise, 
and, hence, so too is the magnetic field induced by the current in the coil.

The switching of the shortcut can be accomplished in a number of ways:
one can heat the shortcut to above its critical temperature, 
as is most often done for large magnets, 
or create a magnetic field higher than the shortcut's critical field\cite{AmeenWiederhold1964,HagedornDullenkopf1974august,
NotoEtAl1995,NotoEtAl1996,GotoEtAl1999},
or even mechanically interrupt it\cite{TsudaEtAl2000,TomitaEtAl2002}.

We do SQUID-based experiments on the mixing chamber stage of a dilution refrigerator cryostat,
which means that aside from being sensitive to magnetic noise,
we are also concerned about heat input.
It is our wish to place a PCS close to our experiment on the mixing chamber stage
such that the wiring between PCS and experiment can be kept short
---
shorter wires are less susceptible to noise and also mechanically less vulnerable
---
while keeping the heat input manageable. 
The magnetic PCS presented in this paper provides us with a solution that suits our needs.

\section{Design and Fabrication}

We set out to design a PCS that:
\begin{enumerate} 
\item Can be operated in a dilution refrigerator, i.e. an environment with very little cooling power on the order of 1 \si{\micro\watt}.
\item Introduces only a low (ideally zero) resistance such that it may be used with small coils of \mbox{$0.1$ mH $-$ $0.1$ H} and still yield a long lifetime $\tau$.
\item Can be charged with a user-adjustable current in the range of $0.01$ A $-$ $1$ A.
\item Has reasonable dimensions, preferably smaller than $5$ cm.
\end{enumerate}
To our knowledge, a PCS that combines all of these features has not yet been constructed.

\begin{figure}[t]
 \centering
   \vspace{-5pt}
   \includegraphics[clip,scale=1]{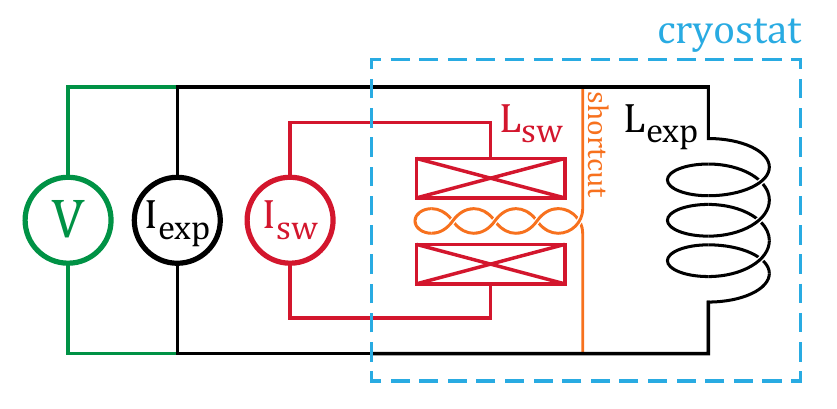}
   \vspace{-15pt}
   \caption{\label{electrical_scheme} In black the basic electrical circuit for making a magnetic field with a coil L\subs{exp} and a current source. 
   Added are a \textcolor{figorange}{superconducting shortcut} 
   that is brought to its resistive state by a 
   \textcolor{figred}{switch coil L\subs{sw}}.
   The response is measured by a \textcolor{figgreen}{voltmeter}.
   The \textcolor{figblue}{blue dashed box} contains the parts that are at low temperature inside a cryostat.}
   \vspace{-5pt}
\end{figure}

In a magnetic PCS we distinguish two coils. 
The experiment coil, which is to be charged with a current $I_{exp}$ and used to perform the experiment of choice, 
and the switch coil with current $I_{sw}$, which is used to bring a superconducting shortcut to the normal state. 
Figure \ref{electrical_scheme} gives a schematic overview.

We fabricate experiment coils from $100$ \si{\micro\metre} diameter NbTi wire with a $13$ \si{\micro\metre} Formvar insulating layer.
We know the inductances of our experiment coils by integrating the voltage response upon charging them:
\begin{equation}\label{Vcharge}
  \int V(t) dt = L_{exp} \Delta I_{exp}.  
\end{equation}
The experiment coils discussed in this paper have inductances $L_{exp} = 60$ mH and $L_{exp} = 440$ \si{\micro\henry}.
For convenience, 
we charge the experiment coil from empty $I_{exp} = 0$ A to $I_{exp} = I_0$ in a single step
and we power up the switch coil as fast as our setup allows for.
The energy dissipated in the shortcut while charging then totals $\frac{1}{2} L_{exp} I_0^2$ as we will derive.

Dissipation can also come from other sources. 
There is Ohmic dissipation in the current lines towards the coils,
which we find to be negligible if care is taken to use superconducting wiring from $4$ K to the mixing chamber stage of the dilution refrigerator 
and if care is taken to thermalize the wiring well.
Further, the rapid change in magnetic field in both the experiment coil and the switch coil may cause dissipation due to eddy currents in normal metals in their vicinity.

The superconducting shortcut
is made of an insulated Niobium wire, $50$ \si{\micro\metre} in diameter.
We twist the wire around itself, such that the mutual inductance to the switch coil is minimized, thereby minimizing the noise input through this channel.
We have chosen to make the shortcut out of Niobium 
because of its high critical current density and critical temperature
and its relatively low critical field.
This means that at the currents we intend to use, 
quasiparticle dynamics are of no concern,
and that tests in liquid Helium are possible.

All superconducting connections are made by spot welding\cite{PhillipEtAl1995} the wires to Niobium sheets of $100$ \si{\micro\metre} thickness. 
In order to get a good superconducting connection 
we strip the wires of their Formvar insulation with a knife and clean them with IPA.
We clean the Niobium sheets by sanding them lightly with sand paper and then wiping them with IPA.
We spot weld the NbTi wires to the Nb sheets with $20-25$ Watt-seconds and the thinner Nb shortcut wire with $8-10$ Watt-seconds.
We have also tried to laser weld the connections, 
but this resulted in higher contact resistances
as well as less mechanical stability.

The switch coil 
should be able to deliver at least $B_{c2} = 400$ mT, the upper critical field of Niobium at $0$ K\cite{FinnemoreEtAl966}.
We construct it from $100$ \si{\micro\metre} diameter Copper clad single core NbTi wire, $62$ \si{\micro\metre} diameter NbTi core, with a $13$ \si{\micro\metre} insulating Formvar layer.
The spindle on which the switch coil is wound is made from PEI and allows for a coil with
an inner diameter of $9$ mm, an outer diameter of $22$ mm and a length of $18$ mm.
In the center of the spindle we leave a hole of $3$ mm diameter through which the shortcut is put.
For sturdiness, we give the spindle $4$ mm thick walls on either side of the coil and apply a layer of Stycast 2850FT to the whole after winding.
We were able to put $N=8672$ windings on the spindle.
The inductance of the switch coil is about $L_{sw} = 0.5$ H.

The NbTi wire leads of the switch coil are twisted and led from 
the mixing chamber stage of the cryostat, where the switch coil is mounted, uninterrupted to the $4$ K stage,
thermalized at the intermediary stages on Copper bobbins.  
From $4$ K to room temperature, the wires are from Copper.
We should avoid sending too high currents,
because these could cause dissipation in the non-superconducting parts of the wiring.
We aimed for a switching current $I_{sw}$ on the order of $1$ A,
at which the switch coil makes a field of
\begin{equation}
  B_{sw} = \frac{\mu_0 N I_{sw}}{l} = 600 \text{mT} ~ > ~ B_{c2} .
\end{equation}

\section{Results and Discussion}

\begin{figure}[t]
 \centering
   \vspace{0pt}
   \includegraphics[clip,scale=1]{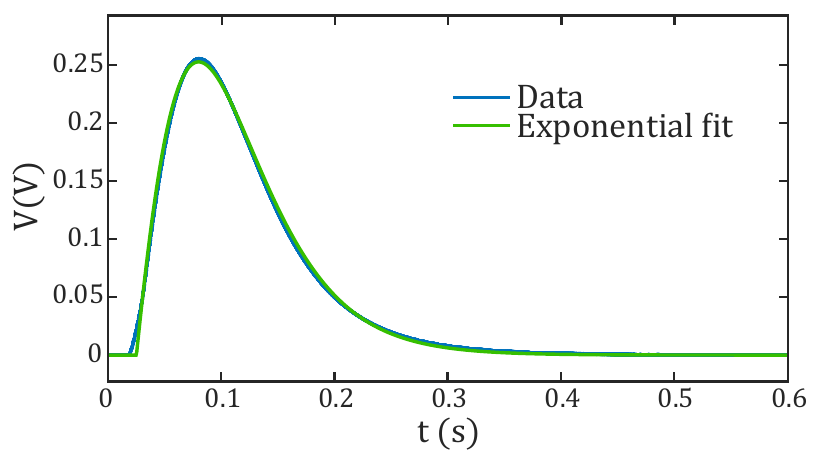}
   \vspace{-10pt}
   \caption{\label{fits} The voltage response of a $60$ mH coil at $T = 11$ mK when charged with $0.5$ A using $I_{sw} = 2$ A.
   The shape of $V(t)$ is determined by the time-dependent normal state resistance $R_{pcs}(t)$ of the shortcut;
   assuming it be exponential, eq. \eqref{Rpcs_t}, we find $R_{ns} = 1.4$ $\Ohm$ and $\tau_{ns} = 73$ ms.}
   \vspace{-5pt}
\end{figure}

We placed the PCS with a $60$ mH experiment coil on the mixing chamber stage of a dilution refrigerator
which reached a minimum temperature of $10.5$ mK.
After some tweaking, we found that we could charge and discharge the experiment coil without dangerously warming up the cryostat by using a switch current of $I_{sw} = 2$ A during $2$ seconds.
Figure \ref{fits} shows a typical voltage response for the case that we charge the experiment coil from $0$ A to $0.5$ A.

If we left the switch activated for longer than $\sim 10$ s, 
we saw a sudden, vigorous increase in temperature.
To avoid having to recondense into the dilution refrigerator every time we switch, 
we therefore limit ourselves to quick switches.
Alternatively, one can choose to charge the experiment coil at a higher temperature when there is more cooling power available 
and cool further down afterwards.
As we will see, the superconducting contacts have such a low resistance that there is ample time before the stored current is appreciably diminished, 
making this a viable possibility.
However, we wanted to verify that it is possible to charge a coil even at the lowest temperature attainable in our cryostat
without introducing a worryingly high heat input.

The shape of $V(t)$ can be understood by assuming that the normal state resistance of the superconducting shortcut grows exponentially to its final value $R_{ns}$ with a time constant $\tau_{ns}$ upon switching\cite{AmeenWiederhold1964}
\begin{equation}\label{Rpcs_t}
  R_{pcs}(t) = R_{ns} ( 1 - e^{-t/\tau_{ns}} ) .
\end{equation}
The values of $R_{ns}$ and $\tau_{ns}$ depend on a multitude of factors 
including the switching field $B_{sw}$ ($\propto I_{sw}$), temperature and cooling power.

The general solution for the PCS's voltage response with time-varying $R_{pcs}(t)$ when charged $I_{exp} = 0\text{ A} \to I_0$ is given by
\begin{equation}\label{PCS_V_t}
  V(t) = I_0 R_{pcs}(t) ~ e^{- \int_0^t \frac{R_{pcs}(t')}{L_{exp}} dt'}.
\end{equation}
Fitting this to the measurement, using the exponential $R_{pcs}(t)$ of eq. \eqref{Rpcs_t}, we find that $R_{ns} = 1.4$ $\Ohm$ and $\tau_{ns} = 73$ ms.

The energy that is dissipated in the shortcut due to (dis)charging of the experiment coil is
\begin{equation}
  E_{diss}(t) = \int^t_0 P_{diss}(t') dt' = \int^t_0 \frac{V^2(t')}{R_{pcs}(t')} dt'.
\end{equation}
For large $t$, it is easy to show that $E_{diss}(t)$ converges to $\frac{1}{2} L_{exp} I_0^2$:
the energy that needs to be dissipated in the shortcut is exactly the energy that is stored in the experiment coil.
This contribution to the dissipation can be made smaller 
by charging the experiment coil in a number of steps rather than in a single step as is done now.
However, in our experiment it is not the dominating factor.

\begin{figure}[t]
 \centering
   \vspace{0pt}
   \includegraphics[clip,scale=1]{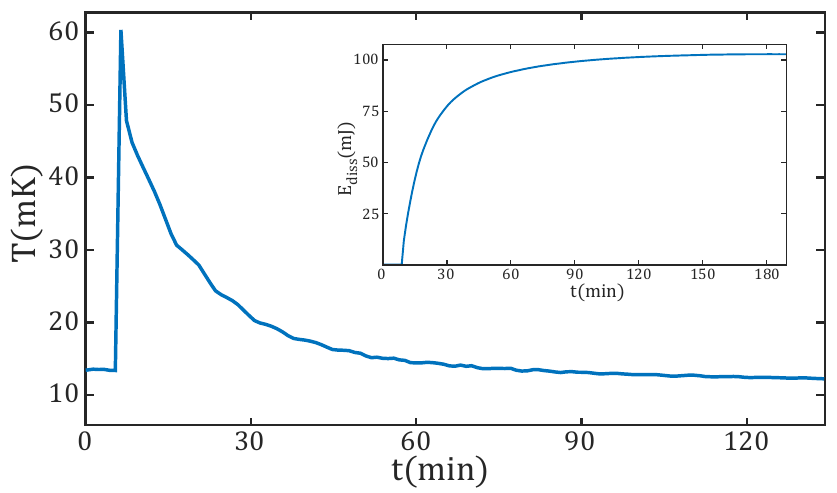}
   \vspace{-10pt}
   \caption{\label{temp_peak_dilfridge} 
   The temperature of the mixing chamber stage versus time, $\Delta t = 1$ minute. 
   When charging the experiment coil with $0.5$ A,
   the temperature increases to $60.5$ mK.   
   The inset shows the estimated dissipated energy in time, which levels off to $103$ mJ.}
   \vspace{-5pt}
\end{figure}

During (dis)charging of the coils, we monitor the temperature of the mixing chamber stage
onto which the PCS is mounted.
Figure \ref{temp_peak_dilfridge} shows the temperature of the mixing chamber stage as a function of time during the switch of figure \ref{fits}.
At the moment of switching, the temperature quickly increases to $60.5$ mK and gradually decreases again afterwards.
Our calibration of the cooling power versus temperature $P_{cool}(T)$ 
allows us to convert the temperature to power dissipation,
and,
consequently, integration yields the dissipated energy in the temperature peak. 
We estimate $E_{diss} \approx 100$ mJ, see the inset in figure \ref{temp_peak_dilfridge}.

Note that this is much more than the energy dissipated in the shortcut, $\frac{1}{2} L_{exp} I_{0}^2 = 7.5$ mJ.
We attribute this to eddy currents:
for the sake of thermalization our coils are securely fastened to the mixing chamber stage of the dilution refrigerator, 
which is a gold-coated copper disk of $1$ cm thickness.
Being a normal metal, induced eddy currents are dissipated here and cause a heating that scales with $(dB/dt)^2$ \cite{Pobell}.
A straightforward way of reducing this heat input
would be to ramp the switch current up and down more gradually 
and to divide the charging of the experiment coil into several small steps.
The eddy currents could be further reduced by keeping the switch and experiment coils away from normal metals or encasing them in a superconducting shield and use normal metal only to provide cooling to the coils.

\begin{figure}[h]
 \centering
   \vspace{5pt}
   \includegraphics[clip,scale=1]{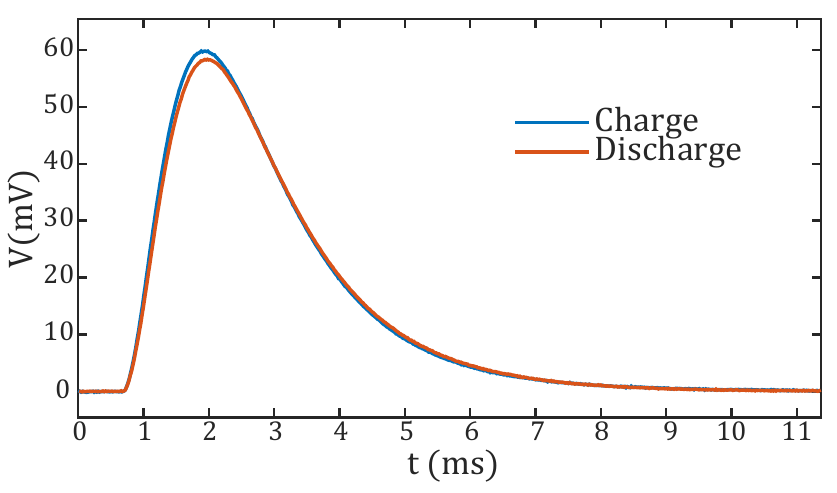}
   \vspace{-10pt}
   \caption{\label{long_4K_meas} 
   The voltage responses at $4$ K when charging to $350$ mA, waiting for a time $t_W = 16.8$ days and then fully discharging.  
   For the discharge voltage response we plot $-V(t)$ to emphasize how little current has been lost.
   We measure the current still flowing in the experiment coil to be $I_W = 346$ mA, 
   therefore $\tau \geqslant 4.2$ years and $R \leqslant 3.3$ p$\Ohm$.
   }
   \vspace{-10pt}
\end{figure}

We measured the quality of the spot welded superconducting joints
by evaluating the residual resistance $R$
in a long-lasting measurement.
Because of the exceptionally low resistance of the joints,
there is no measurable decrease in $I_{exp}$ even after a few days.
We therefore placed a PCS with a $440$ \si{\micro\henry} experiment coil in a vacuum dipstick inserted in a liquid Helium dewar,
charged it from $0$ A to $350$ mA, 
and left it untouched for a little under $17$ days before discharging.
Figure \ref{long_4K_meas} shows the two voltage responses. 
Fitting to equation \eqref{PCS_V_t} gives
$R_{ns} = 375$ m$\Ohm$ and $\tau_{ns} = 1.1$ ms.

The integration of $V(t)$, eq. \eqref{Vcharge}, 
when discharging $I_{exp} = I_W \to 0$ A yields the current still flowing in the experiment coil $I_W$ after waiting a time $t_W$.
It is related to the injected current $I_0$ as
\begin{equation}
  \frac{I_W}{I_0} = e^{-t_W /\tau} , \quad \tau = \frac{L_{exp}}{R} .
\end{equation}
The amount of current still flowing thus allows us to measure the residual resistance in the superconducting circuit $R$.
We measure $I_W = 346$ mA,
which translates to $\tau \geqslant 4.2$ years and $R \leqslant 3.3$ p$\Ohm$.

\section{Conclusions}

We have shown that it is possible to put a $60$ mH coil in the persistent mode carrying a current of $0.5$ A 
in a dilution refrigerator at $11$ mK
using a magnetic Persistent Current Switch.
We find that spot welding all wires to each other via Nb sheets ensures a resistance of less than $3.3$ p$\Ohm$.

\section{Acknowledgements}

The authors wish to thank J.J.T. Wagenaar and M. de Wit for helpful discussions
and G. Koning and F. Schenkel for technical support.
This research is part of the Single Phonon Nanomechanics project of
the Dutch Foundation for Fundamental Research on Matter (FOM).


\bibliographystyle{apsrev}


\end{document}